\documentstyle[aps,preprint]{revtex}
\tightenlines

\newcommand{\be}{\begin{equation}}
\newcommand{\ee}{\end{equation}}
\newcommand{\bea}{\begin{eqnarray}}
\newcommand{\beas}{\begin{eqnarray*}}
\newcommand{\eea}{\end{eqnarray}}
\newcommand{\eeas}{\end{eqnarray*}} 
\newcommand{\ba}{\begin{array}}
\newcommand{\ea}{\end{array}}
\def\ls{\mathrel{\lower4pt\vbox{\lineskip=0pt\baselineskip=0pt
           \hbox{$<$}\hbox{$\sim$}}}}
\def\gs{\mathrel{\lower4pt\vbox{\lineskip=0pt\baselineskip=0pt
           \hbox{$>$}\hbox{$\sim$}}}}

\begin{document}

\draft

\preprint{{TUM-HEP-416/01}}

\title{Final reheating temperature on a single brane }

\author{Rouzbeh Allahverdi~$^{1}$, Anupam Mazumdar~$^{2}$ and
A. P\'erez-Lorenzana~$^{2,3}$}
\address{$^{1}$ Physik Department, TU Muenchen, James Frank
Strasse, D-85748, Garching, Germany. \\
$^{2}$ The Abdus Salam International Centre for Theoretical Physics, I-34100,
Trieste, Italy.\\
$^{3}$ Departamento de F\'{\i}sica,
Centro de Investigaci\'on y de Estudios Avanzados del I.P.N.\\
Apdo. Post. 14-740, 07000, M\'exico, D.F., M\'exico.}


\maketitle

\begin{abstract}
We make a generic remark on thermal history of a single brane cosmology in
models with an infinitely large single extra dimension.  We point out that the
reheat temperature of the Universe is bounded by an excess production of
gravitons from the thermal bath. The actual bound is given by the  brane
tension. If the initial temperature of the Universe is larger than this bound,
then an efficient graviton production shall prevail. However, the brane cools
down gradually as the Kaluza-Klein gravitons take the energy in excess 
away from the brane. The cooling  continues until the radiation dominated 
phase is restored, which occurs before the Big Bang Nucleosynthesis. We 
argue whatsoever be the early evolution of the Universe, the final radiation
dominated phase always starts after the Universe transits from 
non-conventional era to the standard cosmological era. 
\end{abstract}

\vskip60pt


There has been much interest in the possible existence
of a non compact (infinite) extra dimension~\cite{lisa1}.  
This has a striking feature that a four dimensional gravity can be 
thought of as being a zero mode of a $4+1$ dimensional 
anti de-Sitter bulk field which is localized on a hypothetical 
$3+1$ dimensional Poincar\'e invariant brane where we are assumed to live. 
This gravitational zero mode has a unique profile which  decays as we go 
away from the brane due to the presence of a non-trivial warp factor 
which peaks around the  brane. A simple extension of this proposal can 
also solve the hierarchy  between the Planck scale and the electroweak 
scale, if another brane with a negative tension is located  
at a distance where the vacuum expectation value of the
Higgs naturally picks up an electroweak scale due to the presence of 
the warp factor~\cite{lisa2}. It has been noted earlier in 
Ref.~\cite{anupam}, that the thermal history of
such a Universe departs significantly  from the standard lore.
This is mainly due to the fact that the $(0,0)$ component of the 
Einstein's equation contains some new terms which are present due
to the fact that the brane is infinitely thin, and, the matter fields
in the $4+1$ dimensional set-up are actually supposed to be localized on 
such a brane~\cite{binetruy,maeda}. An interesting feature of such a 
brane world system is a continuous mass spectrum of the 
Kaluza-Klein (KK) modes of gravity expanding from zero mass up to 
the Planck scale (for a review, see Ref.~\cite{rubakov}). 
It means, from the five dimensional point of view the gravitons 
can take away any amount of energy in the form of the fifth momentum.
This is in contrast with the usual KK theories 
and all kind of compact extra dimensional models, where the KK 
spectrum is always discrete (see for instance, 
Refs.~\cite{nima,dvali,sarkar1,hall,davidson}). 

\vskip3pt

It is usually believed 
that these KK modes can be excited from a thermal bath once inflation 
comes to an end. As the space is compact they just circulate around 
maintaining their influence on the brane.
Eventhough, these KK modes are weakly coupled to other 
matter fields, they could pose a threat to the synthesis of 
light elements in a similar way as in the case of a massive unstable 
relic particle~\cite{ellis}. 
Although, this is a particular feature in models with compact extra
dimensions, it is not considered to be true in the present case, because
the KK gravitons can truly leave the brane into the extra infinite dimension. 
From the point of view of a four dimensional observer they
notice a loss of energy from the brane. This might be helpful because 
the dangerous KK modes may no longer threat nucleosynthesis whatsoever.

\vskip3pt

Usually, the reheat temperature of the Universe is recognized  
as the largest temperature of the Universe during the radiation epoch that
extends up to the nucleosynthesis era. In almost all the cases the reheat 
temperature of the Universe is constrained in order not to over produce 
weakly coupled particles such as KK gravitons, gravitinos and moduli. 
Especially in the supersymmetric case one has to worry about the 
gravitinos because their mass is $\sim 1$ TeV in gravity mediated 
supersymmetry breaking models, and their coupling to other fields 
is Planck mass suppressed. This is the reason why they decay during
the nucleosynthesis era. When they decay they inject entropy and 
reheat the plasma~\cite{sarkar2}. Very similarly once the KK modes 
are produced they go out of equilibrium from the rest of the thermal 
bath and  their number density redshifts as the Universe 
expands. The mass of these modes remain same when they decay and their
decay products may inject entropy to the Universe before nucleosynthesis,
during nucleosynthesis, and after nucleosynthesis. If they
decay much later, then, they can be  constrained from the diffusion of 
photons in a micro wave background radiation~\cite{davidson,photon}.

\vskip3pt

In this paper we discuss thermal history of a single brane cosmology. 
We argue that the reheat temperature of a
brane is only bounded by a  threshold temperature, known as the {\it normalcy
temperature}  of the KK gravitons. This temperature determines the departure 
from the radiation dominated Universe to a KK dominated Universe. Beyond 
this temperature apart the usual cooling due to the expansion, there
is an extra source of cooling. The brane can also radiate the excess of 
energy to the extra space.  This cooling process becomes less significant
when the temperature of the brane becomes as low as
the normalcy temperature, since below this  temperature the KK
graviton production is not anymore efficient. In what follows we briefly
introduce the cosmology of a single brane. We then  study the abundance of
gravitons and thermal history of the Universe in a more general context.
For consistency we treat gravitons always in terms  of KK modes, although
eventually the five dimensional point of view  helps to understand
their dynamics in the bulk.

\vskip3pt

The cosmology of a single brane is modified significantly due to 
the fact that the energy momentum tensor is localized on a brane with 
$T^{\mu}_{\nu~ brane}=\delta(y)\left(-\rho, p,p,p,0\right)$. 
This  defines an appropriate boundary condition for the 
cosmological parameters in the extra spatial direction and 
changes the Friedmann equation while describing the time 
dependent scale factor on the brane~\cite{binetruy,maeda,csaki,cline}.
In the simplest scenario, where the extra dimension
is supposed to be {\it stable} and free from 
{\it space-like singularities}, 
no time dependent contribution comes from the bulk. In 
particularly,  an extra term appears in the Friedmann equation
which goes as $\sim 1/a^4$, where $a$ is the time dependent
scale factor on the brane~\cite{binetruy}. Such a term is usually 
interpreted as a dark radiation contribution to the brane, and 
actually encodes the information of the time dependence of the 
fifth dimension~\cite{maeda,abdel}. Also, it has been conjectured 
on the basis of AdS/CFT correspondence that such a time dependent 
source term might appear due to the presence of a black hole in 
the bulk (see for instance, Refs~\cite{visser,russell}). However, as it has 
been shown in Ref.~\cite{abdel},  such a term is absent if we assume 
that the bulk is stable. In such a case the Friedmann equation 
simplifies and yields
 \be
 \label{main0}
 H^2=\frac{8\pi}{3M_{\rm p}^2}\rho\left[1+\frac{\rho}{2\lambda}\right]\,,
 \ee
where the brane tension $\lambda$ relates the four dimensional Planck 
mass $M_{\rm p}\approx 10^{19}{\rm GeV}$ to the (fundamental) five dimensional 
Planck scale $M_{5}$ via~\cite{lisa1}
 \be
 \label{main1}
 M_{\rm p}=\sqrt{\frac{3}{4\pi}}\left(\frac{M_{5}^2}{\sqrt{\lambda}}\right)
 M_{5}\,.
 \ee
This is actually a consequence of the cancellation of the  
negative bulk cosmological constant with the brane tension 
$\lambda$~\cite{csaki,cline} that we have assumed. While the 
Friedmann equation is modified, the conservation 
of the energy momentum tensor remains valid on the brane.  
If we demand that successful nucleosynthesis takes place, 
then the second term proportional to $\rho^2$ has to
play a negligible role at a scale $\sim  {\cal O}(\rm{ MeV})$, corresponding
to the era of Big Bang Nucleosynthesis (BBN). Therefore, we have to assume that 
the modified Friedmann equation paves a usual term on the right-hand 
side of Eq.~(\ref{main0}), which is just linear in energy density
\footnote{We will frequently imply Eq.~(\ref{main0}) to be a consequence 
of a non-conventional brane cosmology compared to the standard cosmology 
where $H=\sqrt{8\pi \rho/3M_{\rm p}^2}$.}. This 
naturally leads to constraining the brane tension as 
$\lambda > (1~\rm{MeV})^4$. A more stringent constraint
on the brane tension can be obtained  from 
the validity of the Newtonian gravity in $3+1$ dimensions on length scales 
smaller that $0.2$ mm~\cite{exp}, which leads to constraining the brane 
tension as $\lambda >(1~{\rm TeV})^4$~\cite{dvali2,everett}. 

\vskip3pt

In our case the Universe exits from the non-conventional 
era when the energy density $\rho \sim \lambda$, this happens at 
a temperature $T_{transit} \sim \lambda ^{1/4}$ in a radiation dominated
Universe before BBN.
At energy scales greater than the brane tension the thermal history 
of the Universe can be altered significantly, and some of the 
consequences have already been discussed in Ref.~\cite{anupam}, where 
an upper bound $\lambda \leq (10^{10}{\rm GeV})^4$ 
has been derived. 
This bound  neglects the contribution 
from the KK spectrum, and thus the results of above reference
holds good if the temperature of the
Universe is below the normalcy temperature.  
Once thermalization of the final products of the inflaton has taken place, 
there could be an initial post-inflationary phase which is radiation 
dominated. The graviton production occurs due to thermal processes, such as 
photon-photon fusion, $\gamma+\gamma \rightarrow G_m$, where 
$G_m$ corresponds to the KK graviton of mass $m$. This single 
process is generically Planck mass suppressed~\cite{everett,rizzo},  
such that the cross section goes as  $\sigma_m\sim h^2/M_{\rm p}^2$, 
where $h$ represents the dimensionless coupling constant of the KK mode. 
This can
also be interpreted as the mode number density, which is given by the wave 
function of the graviton along the fifth dimension. The production of the KK
modes occur at all temperatures because   they are distributed continuously on
mass. In the  simple analogy of  a hot radiating plate, one can imagine that
a brane is a hot surface embedded in a cold space,  it radiates 
gravitons with a spectrum which peaks around the brane
temperature. The amplitude of the spectrum  depends on the efficiency of the
graviton production. In order to proceed with our calculation we need to know 
the density of states and for this purpose we have to find out the effective
four dimensional wave function of these modes.

\vskip3pt

The actual calculation for the wave function of the KK graviton 
has been performed strictly in a static limit in 
Ref.~\cite{lisa1,rubakov}. 
The set-up is the following. Let us consider a $5$ dimensional 
anti de-Sitter space (the bulk) where a flat brane of tension 
$\lambda$ is located at $y=0$; here $y$ represents the
infinite fifth dimension. The brane is devoid of any matter $\rho \sim 0$. The
static metric which is a solution to the Einstein's equation of 
this set-up is given by
\begin{eqnarray}
\label{metric}
ds^2=e^{-2\kappa|y|}\eta_{\mu \nu}dx^{\mu}dx^{\nu}+dy^2\,,
\end{eqnarray}
where the constant $\kappa$ in the decaying warp factor 
relates the Planck and the fundamental scales by
\begin{eqnarray}
\label{main2}
\kappa\approx\frac{M_{5}^3}{M_{\rm p}^2}\,.
\end{eqnarray}
The above parameter also plays the role of the effective size 
of the extra dimension, since the KK contribution to gravitational 
interactions on the brane introduces a correction to the Newton's law 
which has a functional behavior of $\sim 1/\kappa r$~\cite{lisa1,dvali2}, 
which is similar to that of one large extra dimension~\cite{nima,dvali}
of size $\kappa^{-1}$. Notice, that the obtained Einstein's solution 
in Eq.~(\ref{metric}), does not hold true if the metric has an arbitrary 
time dependence. The ``static'' KK graviton  wave function in 
Gaussian-normal coordinates at the brane position is then given by
\cite{lisa1,rubakov}
\begin{eqnarray}
\label{wave}
|h_m(y=0)| = {2\over \pi}\sqrt{\kappa\over m} 
	{1\over \sqrt{J_1^2({m\over \kappa}) + N_1^2({m\over \kappa})}}
 \approx \left\{\ba{l l l}
 {\rm const.}\sqrt{\frac{m}{\kappa}} & \mbox{ for } &  m\ll \kappa \\[1em]
 {\rm const.} & \mbox{ for } &  m \gg \kappa 
  \ea \right.
\end{eqnarray}
where $m$ designates the mass of the KK mode (the fifth momentum of the
graviton), and,  ${\rm const.} \sim {\cal O}(1)$. 
We notice,
depending on the mass of the KK mode the projected wave function on the brane
is different. The physical reason behind this effect is the
presence of a volcano potential~\cite{lisa1} felt by the gravitons which
makes all the lighter modes weakly interacting with respect to the modes that
are above the hight of the volcano potential, which is of the order of
$\kappa$. In our case the situation is quite different, since we are certainly
not in a  static solution. We have a matter $\rho \neq 0$ on the brane.
However, we presume  that the structural form of the above equation remains
intact except for some unknown constant factors. In what follows we shall
assume the  sanctity of Eq.~(\ref{wave}) in our analysis.

\vskip3pt

At higher temperatures, $T > \kappa $,  
the total cross section for graviton production 
(summed over all possible final modes) is then given by the sum of both 
the regimes mentioned above in Eq.~(\ref{wave}).
\begin{eqnarray}
\label{cross}
\sigma_{\gamma+\gamma\rightarrow G}\sim \frac{1}{M_{\rm p}^2}\int_{0}^{T} dn
\equiv \frac{1}{M_{\rm p}^2}\int_{0}^{\kappa}\frac{m}{\kappa}\frac{dm}{\kappa}
+ \frac{\rm {const.}}{M_{\rm p}^2}\int_{\kappa}^{T}\frac{dm}{\kappa}
\approx\frac{T}{\kappa M_{\rm p}^2}\,.
\end{eqnarray}
Notice, that in the above expression the main contribution to the cross section
comes from the heavier modes rather than the lighter modes. The cross section 
then  goes linearly in temperature, just as in the case  of a single compact
large extra dimension~\cite{dvali}, where in general the same dependence goes
as $(TR)^\delta$ for $\delta$ extra dimensions compactified on a torus. In fact
in a good approximation the effective number of levels contributing at higher
temperatures is given by $\sim (T/k)$. An interesting point to notice here is
that at low  temperatures, $T < \kappa $, the  cross section goes as 
$\sigma \sim (T^2/\kappa^2 M_{\rm p}^2)\sim T^2/\lambda$,  
which actually mimics the result of two compact large extra
dimensions.  Here we stress that the KK modes are not distributed  uniformly in
energy scales. This is also the reason why the lighter states $ T < \kappa$
induce a  correction to the Newtonian potential as $1/(\kappa r)^2$,  while the
heavier states contribute to the correction as $1/\kappa r$ 
only~\cite{dvali2,everett}. From the five dimensional point of view this
reflects that gravitons with a large fifth momentum are easier to produce
since their energy is above the volcano barrier, whereas the lighter modes have
to cross through such a  barrier, thus their production is less efficient.

\vskip3pt

Now let us consider the evolution of the production of  
KK modes in a simple set-up
by assuming that radiation is dominating the  Universe. 
Our fist goal is to  estimate the largest temperature that the photon bath can
achieve without overproducing gravitons. That is what is known as 
the normalcy temperature. Below this temperature the cooling rate of the
brane due to dissipating energy is less noticeable.
As a first approximation to the problem we
assume that the KK gravitons which have a momentum in the fifth direction
have not gone very far away from the brane. This allows us to count 
their degrees of freedom as if they were lying on the vicinity of the brane. 
Therefore, the equation which governs the individual KK mode number 
density can be given by 
\begin{eqnarray}
\label{evol}
\frac{d n_{G,m}}{dt}+3Hn_{G,m}=\langle \sigma~v\rangle_m ~ 
n^2_{\gamma},
\end{eqnarray}
Notice, that gravitons are actually escaping from the brane, such that 
our present approach actually overestimates their
real number. However, the result will actually 
give us the safest temperature at which the expansion of the three
spatial directions to the brane is not  being  
affected by the KK  gravitons which have being released into the bulk.
In fact, as they propagate at most with the
speed of light on the bulk,  it certainly takes a while for them to go far 
from the brane and relive the brane from their influence. As the 
production is continuous they form a cloud around the brane that 
freely expands into the bulk as the brane cools gradually.
Here we must remember that in Eq.~(\ref{main0}), one has assumed that 
the main contribution to the Hubble expansion comes only from the brane
matter. If a dense cloud of gravitons is surrounding the
brane then their contribution to the expansion  must also 
be taken into account. That may even restore for instance, the ``dark
radiation'' term; $\sim 1/a^4$, where $a$ is the scale factor of the Universe.
We remind the readers that this term has been neglected in Eq.~(\ref{main0}).
Such contributions shall remain negligible as far as the temperature of the
thermal bath on the brane is lower than the normalcy temperature that 
we are about to calculate. By simply assuming the adiabatic expansion
 $a(t)T(t)={\it constant}$, we can in fact simplify Eq.~(\ref{evol}). 
While doing so we may also neglect the evolution
of the individual mode and shall concentrate upon all possible 
KK states excited upto a given temperature. We obtain
\begin{eqnarray}
\label{evol1}
\frac{d(n_{G}/n_{\gamma})}{dT}=-\frac{\langle \sigma ~v\rangle n_{\gamma}}
{H ~T}\,.
\end{eqnarray}
Once the KK states are excited they are no more in 
thermal equilibrium,  we can integrate Eq.~(\ref{evol1}) while assuming
that we are in a standard cosmological era such that 
$H^2\propto \rho/M_{\rm p}$, we get
\begin{eqnarray}
\label{final0}
\frac{n_{G}(T)}{n_{\gamma}}= {\cal D}~\frac{n_{\gamma}(T_{\rm r})
\langle \sigma ~v\rangle}{H(T_{\rm r})}\,,
\end{eqnarray}
where ${\cal D}$ is the dilution factor which depends on the
ratio of the number of relativistic degrees of freedom. In
the Standard Model this ratio can be at most of order 
${\cal D} \sim {\cal O}(10^{-2})$, if the maximum temperature 
is above $\sim 1$ GeV. The temperature $T_{\rm r}$ designates the largest
temperature during radiation era which is also known as the reheat 
temperature of the Universe. We also take $v=1$, henceforth. Now with 
the help of Eq.~(\ref{cross}) and assuming that the relativistic particles
dominate the Universe; $n_{\gamma} \sim T_{\rm r}^3$,
we evaluate the right-hand side of Eq.~(\ref{final0}). The ratio thus 
obtained can not exceed more than one at any later times in order 
to maintain the successes of the nucleosynthesis era and so we obtain
a simple bound on $T_{\rm r}$, which is 
\begin{eqnarray}
\label{final1}
T_{\rm r} \ls T_c \equiv \lambda^{1/4}~= \sqrt{\kappa M_{\rm p}}\,,
\end{eqnarray}
where  $T_c$ represents the {\it normalcy temperature}. 
As mentioned it guarantees that below this temperature
the production rate of gravitons is not efficient enough with 
respect to the number of photons, and  thus, 
the Universe can be safely considered to be in the radiation dominated
phase before BBN. Notice also that $T_c > \kappa$ is actually  
consistent with the assumptions in Eq.~(\ref{cross}). The above 
temperature shall act as a test bed for any departure from the 
radiation era in a standard cosmology.  Further, let us
notice that this temperature is exactly the same as the transit temperature;
$T_{transit}\sim \lambda^{1/4}$, that naively marks the transition  from the
non-conventional era to the standard Universe.  This renders our analysis
completely fool-proof.

\vskip3pt

Let us remark that in the interesting case where one assumes 
$\kappa^{-1}= 0.1~{\rm mm}$, one obtains $T_c \approx 1$ TeV. 
This might render an extreme fine tuning in 
the inflaton coupling to the matter fields to reach  a reheat temperature
which is lower than the normalcy temperature. 
It is worth mentioning that unlike other relics, the KK modes are
being radiated away from the brane to the  infinite extra fifth dimension. 
Therefore, 
the KK modes actually escape all important cosmological constraints 
mainly coming from BBN. Notice, for instance, that unlike 
the case of those four dimensional 
fields with masses around TeV and Planck suppressed couplings, 
which  decay very close to BBN era, 
in the present case a KK mode with the same mass
will be far away from the brane by the time $\tau_{BBN}\sim 1$ sec. 
Therefore, such a mode should have lost
all its interactions to the brane fields, and it will not decay back to the
brane. Therefore, they are totally  harmless. 

\vskip3pt

In order to cross check our previous analysis, let us wonder what happens 
if the Universe prefers to thermalize during the non-conventional era while 
$H \propto (\rho/M_{\rm p}\sqrt{\lambda})$. We can repeat the same
procedure. An important point to mention is that the structural
form of Eq.~(\ref{final0}) shall remain intact in our case~\cite{anupam}.
Now, as $H$ goes like $T^4$, the ratio in Eq. (\ref{final0}) becomes 
temperature independent. This tells 
us that the KK modes have already saturated the photons number density.
This result is consistent with our assumption.

\vskip3pt

It is worth mentioning that the above analysis actually does not preclude the 
possibility of thermalizing the Universe during the non-conventional  era, or,
in general above the normalcy temperature given by Eq.~(\ref{final1}). If the
Universe thermalizes at temperatures larger than the normalcy temperature after
the inflaton has decayed, then the initial radiation  bath might be able to
excite a larger number of KK gravitons overpassing  the photon density. 
This might render the Universe in a phase where the energy stored in the
relativistic species is quickly being released into the bulk in the 
form of KK gravitons, and eventually they decouple at some point from 
thermal history.  
The effective KK number density is then substantially reduced  paving 
a radiation dominated phase, which must be restored at least before 
$\sim {\cal O}(1 {\rm Mev})$. 
At higher temperatures the  effective  number of states   one can excite
follows as $T/\kappa $ for temperatures $T\gg \kappa$, from Eq.~(\ref{wave})
and Eq.~(\ref{cross}).

\vskip3pt

Here we make an important remark on the difference between the cosmological
models of a single brane with that of a compact large extra dimensions. In the
compact large extra dimensions it has been noticed that the normalcy
temperature of the Universe has to be $\leq (1-100)$ MeV in order not to over
produce the KK gravitons if the fundamental scale which is $4+2$ dimensional
gravitational constant is  $\sim {\rm TeV}$~\cite{nima,dvali}. In this case
eventhough the produced gravitons have a momentum along the bulk, but 
the bulk is a compact space which allows these modes to hit the 
brane more frequently eventhough the size of the compact dimensions are as
larger as millimeters. This means that these KK modes do not 
decouple from the brane, but their presence is felt physically on the  
brane. Unlike the infinite dimension case these modes in compact 
extra dimensions eventually decay on the brane matter which puts 
severe constraints on their number density. 
It has been shown that the reheat temperature of the Universe 
must be lower than the normalcy temperature. In order to achieve 
this one must promote the inflaton field as a bulk field, whose decay products 
reheats the Universe. It has also been noticed that in order to 
provide a dynamical mechanism to stabilize  the extra compact 
dimensions and to inflate the $3+1$ spatial  
dimensions which could also provide the right amount of the observed
density perturbations in the Universe, one needs to promote the inflaton field
to the higher dimensions~\cite{abdel1,abdel2}. This inevitably leads to the
Planck mass suppressed couplings between the  inflaton and the Standard Model
fields which resides only in our world. As an important consequence of this the
reheat temperature of the Universe is always below the normalcy temperature and
it is roughly given by $\sim (1-10)$ MeV. This is precisely the reason why such
a low reheat temperature is able to prohibit any possibility of a KK
domination  just before nucleosynthesis, eventhough the temperature of the
Universe at that time is larger than the mass gap of the discrete KK states
which is given by an inversely proportional to the size of the extra 
dimension. 
However,  in the case of a single brane  the situation
is quite different. First of all there is no compelling reason why 
the Universe must thermalize
into a radiation bath below the normalcy temperature $T_c$ defined in
Eq.~(\ref{final1}), especially since the inflaton field may reside 
on the brane itself \cite{maartens}. This leads to a natural question of 
what actually happens if the Universe prefers to thermalize above this 
temperature. This is the topic we shall study next.

\vskip3pt

It is quite evident, that if the KK modes are produced in such a way that their
number density overshoots the other relativistic species, then the evolution of
the Universe would not be that of the radiation era and  the radiation
domination could not be recovered until the last of the KK  modes in excess 
has leaked away far from the brane. This shall be regarded as a
{\it final reheating temperature} denoted by $T_{final}$. A naive
approach tells us that this occurs when the wavefunctions of the gravitons
which is still moving away from the brane has at least moved a distance 
equivalent to the inverse mass of the last possible mode in
excess. This distance then corresponds to the KK mass of order 
$\kappa$, since all the modes above this mass contributes to the effective 
number of relativistic degrees of freedom. Thus, $m \sim \kappa$ 
defines our physical mass of the KK mode. Notice, 
it also means that the cloud formed due to these gravitons have moved away 
a distance equivalent to the effective size of the fifth dimension, 
such that the gravitational interactions to the brane fields become
subleading. As the cloud can expand at most at the speed of light 
since gravitons are actually massless in five dimensions, the decoupling 
from the brane matter takes place at a time scale given by
\begin{equation}
\tau_{dec}\sim \kappa^{-1}\,.
\end{equation}
When this happens the
Universe comes back to a radiation dominated with a standard cosmology. 
Therefore, we estimate the final reheating temperature by equating 
$H\sim T^2/M_{\rm P}$ and $\kappa$. It is interesting to note 
that one obtains exactly the same value as the normalcy temperature,
\be 
T_{final} = \lambda^{1/4}~\,.
\ee
This we recognise as the largest temperature a brane world can have 
in the radiation dominated era if the Universe thermalizes before
the normalcy temperature.

\vskip3pt

An alternative explanation can be illustrated by calculating the cooling 
rate of the relativistic thermal bath. As the KK mode leaves the brane,
the brane looses the energy at a rate given by~\cite{russell};
\be 
 {\dot \rho_\gamma\over \rho_\gamma}= -
 {\langle \sigma E\rangle n_\gamma^2\over \rho_\gamma}
 \approx - C {\rho_\gamma\over \kappa M_{\rm P}^2}~,
 \ee
where the dimensionless 
constant factor $C\sim 0.1$ is given in terms of the distribution functions
of the fields of the thermal bath~\cite{russell}. 
One can easily integrate the above equation to obtain the time scale 
when the brane density has dropped down to $\rho_\gamma = \lambda$.
One obtains the decoupling time scale upto an order one numerical factor, 
\be
\tau_{dec} \sim {\kappa M_{\rm P}^2 \over \lambda} = \kappa^{-1}~.
\ee
This is a familiar result which we have obtained earlier.

\vskip3pt

We notice that the cloud of gravitons that surrounds the brane 
may modify the thermal history of the brane.
Indeed, when the density of gravitons around the brane is 
not negligible, then the  Hubble expansion  becomes a function of the 
fifth dimension~\cite{binetruy}
\be 
H^2(y) \approx  {(\rho_B - \Lambda) \over 6M_5^3} + 
\left({a'\over a}\right)^2~,
\label{hubble2}
\ee
Where $\Lambda$ is the bulk cosmological constant, $\rho_B$ is the bulk
matter density and prime denotes the derivative with respect to $y$. 
At the position of the brane the last term in 
Eq.~(\ref{hubble2}) along with  $\Lambda$ term  reduces to a
$\rho$ squared contribution, where $\rho$ is matter density on 
the brane as depicted in Eq (\ref{main0}). In order to understand the 
dynamics, one requires a profile of the  graviton density 
within the cloud. Besides this, it has also
been argued that the energy which has been radiated  
into the bulk might form a black hole. 
Such a scenario if happens introduces an extra contribution to the expansion 
which goes as $\sim 1/a^4$~\cite{rubakov,russell}. 
As Ref.~\cite{russell} argues, the presence or formation  of a black hole 
in the bulk is important only when $T\sim M_5$.
To avoid such a $\sim 1/a^4$ contribution
the maximal temperature of the initial thermal bath
on the brane should be smaller than $M_5$, which is the  
natural cut-off. However, we are much below this temperature and 
under this assumption we may justify the above expression Eq.~(\ref{hubble2}).

\vskip3pt

In order to have a very rough estimation of the expansion due to
the presence of the cluod, we assume that the 
energy density of the Universe is governed by the bulk density
of relativistic  modes. This is equivalent to the assumption of 
demanding $H^2\approx \rho_B/M_5^3$ in Eq.~(\ref{hubble2}). 
The temperature of such a bath is obviously larger than the normalcy 
temperature given by Eq.~(\ref{final1}).
To a good approximation we may consider the five dimensional 
graviton density to be uniform within distance $\kappa^{-1}$. 
This sets up a scale for the graviton cloud, which determines the
bulk energy density as $\rho_B \approx \kappa \rho_{KK}$, where 
$\rho_{KK}$ is an effective four dimensional KK density. 
This gives an uniform expansion rate parallel the brane, which reads
\be 
 H^2\approx {\rho_{KK}\over M_{\rm P}^2}~.
\ee
From the effective four dimensional point of view 
the situation mimics that in Ref.~\cite{sarkar1}.
The temperature dependence of the effective density is determined by
\begin{eqnarray}
\label{rhokk}
\rho_{KK} \approx \left(\frac{T}{\kappa}\right)T^4\,.
\end{eqnarray}
The term in a bracket corresponds to
the relativistic degrees of freedom which is the number of
KK modes that can be estimated from the wave function Eq.~(\ref{wave}).
Notice, from the five dimensional point of
view this only tells us that $\rho_B\sim T^5$, as it should be expected on
purely dimensional grounds. Given this, the Hubble parameter of the 
parallel directions to the brane is now modified to 
\begin{eqnarray}
\label{new}
 H(T) \sim \frac{T^{5/2}}{\kappa^{1/2} M_{\rm p}}\,.
\end{eqnarray}
Let us point out  that the rate of expansions is actually faster 
compared to the standard behavior $H \propto T^2/M_{\rm p}$. This 
can also be understood from purely  five dimensional point of 
view by inspecting Eq.~(\ref{hubble2}). We notice that the bulk
energy density $\rho_B$ is only suppressed by the fundamental scale rather
than the Planck scale, which is an outcome of a single brane set-up.

\vskip3pt

Let us conclude with some remarks. The evolution of the early Universe 
in a single brane cosmology could be quite  different than naive
expectations. In order to solve homogeneity and the flatness problem one
requires a phase of inflation in these models. Inflation might occur in the
non-conventional era \cite{maartens}, or, in a conventional era. Depending on
this initial phase the Universe might thermalize in a different way. The
thermalization process  inevitably renders the Universe as a radiation 
dominated era, either in a non-conventional era, or, in a standard era. 
In the former case the Universe undergoes inevitably through a KK 
graviton overproduction phase.  During this period the energy of 
the photon bath is quickly transfered into the bulk in
the form of  gravitons which leave the brane once they are produced. 
This can be recognized as a cooling of a brane, which comes to an end  
only when the temperature of the radiation bath has 
dropped down the normalcy temperature.  This happens  
well after the standard $\rho$ behavior of $H^2$ has been
established. The transition to the standard cosmology takes place 
when the last of the KK modes in excess has been dissipated from the 
brane a distance larger than the effective size of the fifth dimension
$\kappa^{-1}$. This estimates the final reheat temperature 
$T_{r} \approx  T_{c} \sim \lambda^{1/4}$.
If thermalization occurs already in the standard cosmology, the
initial radiation dominated phase shall be maintained.
We conclude by saying that the temperature associated with this
threshold can be safely thought as being the largest 
temperature of the Universe in a radiation era. 
It is worth noticing that since the KK modes escape into
the bulk, they do not posse any serious problem for nucleosynthesis.

\vskip10pt

 {\it Acknowledgements.}   
The authors would like to thank the  referee and Kari Enqvist
for helpful comments.
A.M. acknowledges the hospitality of the Helsinki Institute of Physics
where part of the work has been carried out. The work of R.A. is supported by
``Sonder-forchschungsberich 375 f$\ddot{\rm u}$r Astro-Teilchenphysik''
der Deutschen Forschungsgemeinschaft. A.M. acknowledges the
support of {\bf The Early Universe Network} HPRN-CT-2000-00152.


\end{document}